\documentclass[11pt]{article}
\newcommand{\comm}[1]{}
\def\forarXiv{\comm}\def\forarXiv{}
\usepackage{amssymb}
\usepackage{amsmath}\usepackage{graphicx} \usepackage[numbers]{natbib}
\def\citet{\cite}

\usepackage{times}\hfuzz=10pt 

\newtheorem{theorem}{Theorem}
\newtheorem{lemma}{Lemma}
\newtheorem{corollary}{Corollary}
\newtheorem{remark}{Remark}

\def\defi{\stackrel{{\scriptscriptstyle \Delta}}{=}}

\def\o{\omega}
\def\O{\Omega}

\def\F{{\cal F}}

\def\R{{\bf R}}

\def\Z{{\cal Z}}
\def\ZZ{{\bf Z}}

\def\H{{\cal H}}

\def\C{{\bf C}}

\def\T{{\mathbb{T}}}

\newcommand{\be}{\begin{equation}}
\newcommand{\ee}{\end{equation}}
\newcommand{\bd}{\begin{displaymath}}
\newcommand{\ed}{\end{displaymath}}
\newcommand{\ba}{\begin{array}{ll}}
\newcommand{\ea}{\end{array}}
\newcommand{\baa}{\begin{eqnarray}}
\newcommand{\eaa}{\end{eqnarray}}
\newcommand{\baaa}{\begin{eqnarray*}}
\newcommand{\eaaa}{\end{eqnarray*}}

\def\ew{\left(e^{i\o}\right)}
\def\BL{{\scriptscriptstyle BL}}
\def\BLO{L_{2}^{\BL,\O}(\R)}
\def\T{{\mathbb{T}}}
\def\ZZ{{\mathbb{Z}}}

\def\TT{{\cal T}}
\def\XNN{\ell_2^\BL}
\def\BNN{{\mathbb{L}}^W(\T)}
\date{Submitted March 14, 2016 }
\title{ On Nyquist-Shannon Theorem with one-sided half of sampling sequence }
\author{
Nikolai Dokuchaev}
 
\begin{document}
 \vspace{-0.5cm}      \maketitle
\def\brea{}
\def\breakk{}
\def\brea{\nonumber\\ }
\def\breakk{\nonumber\\&&}
\def\break{}
\def\break{\nonumber\\ }
\begin{abstract} The
 classical sampling  Nyquist-Shannon-Kotelnikov theorem states that
 a band-limited continuous time function can be uniquely recovered without error  from a infinite two-sided sampling series taken with a sufficient frequency.  This short note shows that the function can be recovered from any one-sided semi-infinite half of any oversampling series, with the same boundary for admissible frequencies as in the classical theorem.
\index{in
IEEE:}\let\thefootnote\relax\footnote{The author is with  Department of
Mathematics and Statistics, Curtin University, GPO Box U1987, Perth,
Western Australia, 6845 (email N.Dokuchaev@curtin.edu.au).  \index{This work  was
supported by ARC grant of Australia DP120100928 to the author.}}
\end{abstract}\section{Introduction}
This short note suggest a modification of the
 classical sampling theorem that is also known as
Nyquist-Shannon theorem, Nyquist-Shannon-Kotelnikov theorem,   Whittaker-Shannon-Kotelnikov theorem, Whittaker-Nyquist-Kotelnikov-Shannon theorem,
which is one of the most basic results in the theory of signal processing and information
science. This  theorem states that
 any band-limited function can be uniquely recovered without error  from a infinite two-sided equidistand
sampling sequence  taken with sufficient frequency.   This result  was obtained independently by four  authors  \cite{Whit,Nyq,Kotel,Shannon}.
Numerous  extensions of this theorem were obtained,
 including
the case of nonuniform sampling and restoration of the signal with mixed samples;
see some recent literature review in \cite{jerry,U50,V70}.
In particular, it is known  that a bandlimited function can be recovered without error from a sample if a finite number of sample values is unknown. It is also known  \cite{F95}
that the function can be recovered without error  if a infinite subsequence
of the sample values located on the same distance  is missing, with some additional constraints on the
signal band and frequencies \cite{F95}. In this paper, we show that, with the same boundary for admissible frequencies as for the classical Nyquist-Shannon-Kotelnikov Theorem,  any band-limited function can be uniquely recovered without error  from any  one-sided semi-infinite
 half of any oversampling equidistant sampling series, with the same boundary for admissible frequencies as in the classical theorem. This means that  any one-sided semi-infinite half of  equidistand oversampling series  can be deemed redundant: the function still can be restored without error from the remaining part.
\section{Some definitions}\label{secMain}
We denote by $L_2(D)$ the usual Hilbert space of complex valued
square integrable functions $x:D\to\C$, where $D$ is a domain.

For $x(\cdot)\in  L_2(\R)$, we denote by $X=\F x$ the function
defined on $i\R$ as the Fourier transform of $x(\cdot)$;
$$X(i\o)=(\F x)(i\o)= \int_{-\infty}^{\infty}e^{-i\o t}x(t)dt,\quad
\o\in\R.$$ Here $i=\sqrt{-1}$ is the imaginary unit. For $x(\cdot)\in L_2(\R)$, the
Fourier transform $X$ is defined as an element of $L_2(i\R)$, i.e.
 $X(i\cdot)\in L_2(\R)$).

For $\O>0$, let $\BLO$ be the subset of $L_2(\R)$ consisting of
functions $x$  such that $x(t)=(\F^{-1}
X)(t)$, where $X(i\o)\in L_2(i\R)$ and  $X(i\o)=0\ \hbox{for}\ |\o|>\O\}$.

We denote by $\ZZ$  the set of all integers. \par
\section{The main result}

\begin{theorem}\label{ThM} Let $\O>0$ and  $\tau\in (0,\pi/\O)$ be given. Let
$\{t_k\}_{k\in\ZZ}\subset\R$ be a sequence such that $t_k-t_{k-1}=\tau$ for all $k$. For any $s\in\ZZ$, a band-limited function
$f\in \BLO$ is uniquely defined by the values $\{f(t_k)\}_{k\le s}$.
\end{theorem}
\begin{remark}\label{rem1} The value $\tau=\pi/\O$  is excluded in Theorem \ref{ThM}, meaning that the series $\{t_k\}$ oversamples $f$; this is essential for the proof. This value is allowed in the classical Nyquist-Shannon-Kotelnikov Theorem with two-sided sampling series that states that $f\in \BLO$ is uniquely defined by the values $\{f(t_k)\}_{k\in\ZZ}$
if $\tau\in (0,\O/\pi]$.
\end{remark}
\begin{remark}
Theorem \ref{ThM} considers the left hand half  $\{f(t_k)\}_{k\le s}$ of the sampling series; it is
convenient  for representation of  past historical observations, for instance,  for predicting problems. However,  the same statement  can be formulated
for the right hand half  $\{f(t_k)\}_{k\ge s}$ of the sampling series.
\end{remark}

\begin{corollary}\label{corr1}
Theorem \ref{ThM} implies that, for any finite set $S$,
$f$ is uniquely defined by the values   $\{f(t_k)\}_{k\in\ZZ\backslash S}$.
\end{corollary}

The  fact that, for any finite set $S$,
$f\in\BLO$ is uniquely defined by the values $\{f(t_k)\}_{k\in\ZZ\backslash S}$, is known; it
was established in \cite{F91} by a different method.  Theorem \ref{ThM} extents this result: it
shows that the same is also true
for infinite sets
$S=\{t:\ t> s\}$, for any given $s\in\ZZ$. It is known that the same is not true for some other infinite sets. For example,
if $S=\{t_{2k+1},\ k\in\ZZ\}$ and $2\tau>\O/\pi$, then
$f\in \BLO$ is not uniquely defined by the values $\{f(t_k)\}_{k\in\ZZ\backslash S}$, since the frequency of the sample
 $\{f(t_{2k})\}_{k\in\ZZ}$ is lower than is required by the Nyquist-Shannon-Kotelnikov Theorem; see more detailed analysis in \cite{F95}.

\section{Proofs}  It suffices to proof Theorem \ref{ThM} for
 $s=0$ only; the extension on $s\neq 0$ is straightforward.

Let us introduce some additional notations first.

   We denote by $\ell_2$ the set of all
sequences $x=\{x(k)\}_{k\in\ZZ}\subset\C$, such that
$\sum_{k=-\infty}^{\infty}|x(k)|^2<+\infty$.
  We denote by $\ell_2(-\infty,0)$ the set of all
sequences $x=\{x(k)\}_{k\le 0}\subset\C$, such that
$\sum_{k=-\infty}^{0}|x(k)|^2<+\infty$.

\par Let $\T=\{z\in\C:\  |z|=1\}$.
\par
For  $x\in \ell_2$, we denote by $X=\Z x$ the
Z-transform  \baaa X(z)=\sum_{k=-\infty}^{\infty}x(k)z^{-k},\quad
z\in\T. \eaaa Respectively, the inverse Z-transform  $x=\Z^{-1}X$ is
defined as \baaa x(k)=\frac{1}{2\pi}\int_{-\pi}^\pi
X\left(e^{i\o}\right) e^{i\o k}d\o, \quad k=0,\pm 1,\pm 2,....\eaaa
For $x\in \ell_2$, the trace $X|_\T$ is defined as an element of
$L_2(\T)$.

\par
For $W\in (0,\pi)$, let $\BNN$ be the set of all mappings $X:\T\to\C$ such
that $X\ew \in L_2(-\pi,\pi)$ and $X\ew =0$ for $|\o|>W$. We will call the the corresponding processes $x=\Z^{-1}X$
{\em band-limited}.

Consider the Hilbert spaces of sequences $\ell_2$ and
$\ell_2(-\infty,0)$.

Let $\ell_2^\BL$ be the subset of $\ell_2(-\infty,0)$ consisting of sequences
$\{x(k)\}_{k\in \ZZ}$ such that $x=\Z^{-1}X$ for some $X\ew \in \cup_{W\in (0,\pi)}\BNN$.
 Let $\ell_2^\BL(-\infty,0)$ be the subset of $\ell_2(-\infty,0)$ consisting of traces
$\{x(k)\}_{k\le 0}$ of all  $x\in\ell_2^\BL$.

\begin{lemma}\label{propU} For
  any $x\in\XNN $, there exists an unique
unique  $X\in\cup_{W\in(0,\pi)}\BNN$ such that $x(k)=(\Z^{-1} X)(k)$ for $k\le 0$.
\end{lemma}
\par
By Lemma \ref{propU}, the future $\{x(k)\}_{k>0}$ of a
band-limited process $x=\Z^{-1}X$, $X\in\BNN$,
 is uniquely defined by its  history
$\{x(k),\ k\le 0\}$.  This statement represent a reformulation in the deterministic setting
of  the classical Szeg\"o-Kolmogorov Theorem for stationary Gaussian processes \citet{K,Sz,Sz1,V}.
\par
{\em Proof of Lemma \ref{propU}}.  The proof  follows from predictability results for band-limited discrete time processes obtained  in \cite{D12a,D12b}. For completeness, we will provide a direct and independent proof. \forarXiv{ (This proof can be found in \cite{Df}).}
Let $D\defi\{z\in\C: |z|< 1\}$. Let  $H^2(D^c)$ be the Hardy space of functions that are holomorphic on
$D^c$ with finite norm
$\|h\|_{\H^2(D^c)}=\sup_{\rho>1}\|h(\rho e^{i\o})\|_{L_2(-\pi,\pi)}$.
   In this case,
$\TT=\{t:\ t\le 0\}$.
 It suffices to
prove that if $x(\cdot)\in\XNN $ is such that $x(k)=0$ for
$\le 0$,  then either  $x(k)=0$ for $k>0$ or  $x\notin\ell_2^\BL$.
If  $x(k)=0$ for $k>0$, then   $X=\Z x\in H^2(D^c)$. Hence, by the property of the Hardy space,
 $X\notin\cup_{W\in(0,\pi)}\BNN$; see e.g. Theorem 17.18 from \cite{Rudin}. This
completes the proof of Lemma \ref{propU}.

We are now in the position to prove Theorem \ref{ThM}. Consider
a sequence of samples
\baaa x(k)=f(t_k)=\frac{1}{2\pi}\int_{-\O}^\O
F\left(i\o\right) e^{i\o t_k}d\o, \quad k=0,\pm 1,\pm 2,....
\eaaa
Since $t_k=k\tau$, we have that
\baaa
x(k)=\frac{1}{2\pi}\int_{-\O}^\O
F\left(i\o\right) e^{i\o \tau k}d\o=\frac{1}{2\pi \tau}\int_{-\tau\O}^{\tau\O}
F\left(i\nu/\tau\right) e^{i\nu k}d\nu\break=\frac{1}{2\pi}\int_{-\tau\O}^{\tau\O}
G\left(e^{i\nu}\right) e^{i\nu k}d\nu.
\eaaa
Here $G$ is such that $G\left(e^{i\nu}\right)=\tau^{-1}F(i\nu/\tau)$.
We used here a change of variables $\nu=\o\tau$.  Since
$F\left(e^{i\nu/\tau}\right)\in L_2(i\R)$, it follows that $G\left(e^{i\nu}\right)\in L_2(\T)$.
By the assumption that $\tau<\pi/\O$,  it follows that $\tau\O<\pi$ and $x\in\ell_2^\BL$.
By the Nyquist-Shannon-Kotelnikov Theorem,
it follows that the function $f$ is uniquely defined by
the two-sided sequence $\{x(k)\}_{k\in\ZZ}=\{f(t_k)\}_{k\in\ZZ}$.
Further, Lemma \ref{propU} implies  that a sequence $x\in\ell_2^{\BL}$ is uniquely defined by
its trace  $\{x(k)\}_{k\le 0}$.
 This completes the proof of  Theorem \ref{ThM}. $\Box$
\section{Discussion and future developments}
\begin{enumerate}
\item To apply the classical Nyquist-Shannon-Kotelnikov Theorem for the data recovery, one has to
restore the Fourier transform $F=\F f$ from the two-sided sampling series $\{f(t_k)\}_{k\in\ZZ}$. This procedure is relatively straightforward. In contrast, application of Theorem \ref{ThM} for the data recovery requires to restore Z-transform $G\left(e^{i\nu}\right)=F(i\o/\tau)$ from an one-sided half of the sampling series. By Lemma \ref{propU}, this task is feasible; however, it is numerically challenging.  Some numerical algorithms based on projection were suggested in \forarXiv{\cite{Df} and} \cite{D13}.
\item
Some infinite equidistant sets of sampling
points that can be redundant for recoverability of  the underlying function were described in \cite{F95}. It could be interesting to find other infinite sets with this feature.
\item
It could be interesting to investigate if recovery of $f$ suggested in  Theorem \ref{ThM} is robust with respect to errors in location of the sampling points  $t_k$.
\item It is unclear if our approach based on predictability of discrete time processes is  applicable to processes defined on multidimensional lattices.
It could be interesting to extend this approach on process $f(t)$, $t\in\R^2$, using the setting from \cite{PM}.
\end{enumerate}
{\subsection*{Acknowledgment} This work  was
supported by ARC grant of Australia DP120100928 to the author.}

\end{document}